\documentclass[preprint]{aastex61}

\received{}

\shorttitle{Observations of HDO}
\shortauthors{Kulczak}

\begin{document}

\title{Analysis of Low Excitation HDO Transitions Toward the High-Mass Star-forming Regions
                            G34.26+0.15, W51e$_{1}$/e$_{2}$, and W49N}

\author{Magda ~Kulczak-Jastrz\c{e}bska}

\affiliation{Astronomical Observatory of the Jagiellonian University, 30-244 Orla 171,
Krak\'ow, Poland; kulczak@oa.uj.edu.pl}

\begin{abstract}

   {We present observations of the ground state
    1$_{0,1}$--0$_{0,0}$ rotational transition of HDO at
    464.925 GHz and the 1$_{1,0}$--1$_{0,1}$ transition at 509.292 GHz
    towards the three high-mass star forming regions: G34.26+0.15, W49N
    and W51e$_{1}$/e$_{2}$, carried out with the
    Caltech Submillimeter Observatory. The latter transition is
    observed for the first time from the ground. The spectra are
    modeled, together with observations of higher-energy HDO
    transitions, as well as submillimeter dust continuum fluxes from
    the literature, using a spherically symmetric radiative transfer
    model to derive the radial distribution of the HDO
    abundance in the target sources. The abundance profile is divided
    into an inner hot core region, with kinetic temperatures higher
    than 100~K and a cold outer envelope with lower kinetic
    temperatures. The derived HDO abundance with respect to
    H$_2$ is (0.3--3.7)$\times 10^{-8}$ in the hot inner region ($T > 100
    \mathrm{K}$) and (7.0--10.0)$\times 10^{-11}$ in the cold outer envelope. We also used two
    H$_{2}^{18}$O fundamental transitions to constrain the H$_{2}$O abundances in the outer envelopes.
    The HDO/H$_{2}$O ratios in these cold regions are found to be (1.8--3.1)$\times 10^{-3}$ 
    and are consequently higher than in the hot inner regions of these sources.}
\end{abstract}
   
%
\section{Introduction}

During the cold phase preceding the formation of stellar objects,
molecules freeze-out onto dust grains, forming H$_2$O-
dominated ice mantles, mixed with other less-abundant species. The
low temperature and the disappearance of most molecules, especially
CO, from the gas phase trigger a peculiar chemistry leading
to high abundances of deuterated species. Molecules tend to attach a D
atom rather than an H atom, because deuterated species have larger
reduced masses and consequently lower binding energies, arising from
lower zero-point vibrational energies. Ion-molecule reactions in the
gas phase \citep{Brown89} and reactions on the grain surfaces
\citep{Tielens83} are the two possible mechanisms responsible for
deuterium enrichments in heavy molecules. The reactions involved are
exothermic, which is why significant deuteration levels can be
expected in the cold ISM. In the warmer phase, only very little
fractionation is expected to occur, because the energy barrier could
be overcome by the elevated temperature. However, at the temperatures
deduced for the hot core regions (100--200~K) the ice mantles evaporate and
the gas again becomes enriched in deuterated species, with abundances
elevated compared to the cosmic D/H ratio for a short period, before
the chemistry reaches steady-state. These enhancements in the hot cores
reflect, to some degree, the grain mantle composition in the earlier,
colder cloud phases.
Although the processes leading to the water deuteration are not fully
understood, they are clearly related to the grain-surface chemistry
and the observed HDO/H$_{2}$O ratio reveals the chemical and physical
history of the protostellar materials \citep{Cazaux11, Codella10}. A
recent review of water chemistry can be found in \cite{vanDishoeck13}
and a review of the latest observational results from \emph{Herschel}.\\

Early studies of water deuterium fractionation in high-mass hot
  cores were performed almost twenty years ago, when the submillimeter
  spectrum was largely inaccessible \citep{Jacq90,
  Schulz91, Gensheimer96, Helmich96}. The HDO abundance has been
  recently determined toward the high-mass hot core G34.26+0.15 and  the intermediate-mass protostar NGC7129 FIRS2
  \citep{Fuente12, Liu13, Coutens14}. Here, we present the new observations
  of the ground state rotational transition of HDO at
  464.925 GHz and the first excited transition at 509.292 GHz toward
  three high-mass star forming regions: G34.26+0.15,
  W51e$_{1}$/e$_{2}$ and W49N. The combination of data taken with the same telescope 
of both the ground state transition (1$_{0,1}$--0$_{0,0}$)
and the first excited line (1$_{1,0}$--1$_{0,1}$) provide better constraints on the source structure.
 These HDO transitions are studied for the first time to probe the structure 
of the  envelope of the  W51e$_{1}$/e$_{2}$ and W49N hot cores. In this paper, 
 we aim at determining  the HDO
  fractional abundances relative to H$_2$ in the inner and
  outer region of the core in our target sources using, the static
  radiative transfer code of \citet{Zmuidzinas95}.
  We also used two H$_2^{18}$O fundamental transitions 
  observed by \citet{Flagey13} to constrain the H$_2$O abundance and the
  HDO/H$_2$O  ratio in the outer envelopes.


\section{Observations}
Water is difficult to study from the ground, due to its strong
presence in the Earth's atmosphere. However, many HDO lines,
including the ground state 1$_{0,1}$--0$_{0,0}$ rotational transition
and the 1$_{1,0}$--1$_{0,1}$ excited transition studied here, lie in
atmospheric windows, where observations are possible from high 
sites, under good weather conditions. Table~1 lists the HDO
transitions included in the present study
and the H$_2^{18}$O transitions which we use to constrain 
the HDO/H$_2$O ratio in the outer envelope.

\begin{table}
\begin{center}
\caption{Observed and modeled HDO and H$_2^{18}$O rotational transitions.}
\label{table:1}
\begin{tabular}{lll cccccc}
\tableline\tableline
 Species &Transition & Frequency  & $E_\mathrm{up}$ & $A_\mathrm{ul}$ & Telescope & FWHM \\
         &           &  (GHz)     &      (K)        &    (s$^{-1}$)   &      & (\arcsec)  \\      
\tableline
 HDO &1$_{0,1}$--0$_{0,0}$ & 464.925 & 22.3 & $1.69 \times 10^{-4}$ &CSO& 15.7 \\
     &1$_{1,0}$--1$_{0,1}$ & 509.292 & 47.0  & $2.30 \times 10^{-3}$ & CSO & 15.7 \\ 
     &2$_{1,1}$--2$_{1,2}$ & 241.562 & 95.2 & $1.18 \times 10^{-5}$ & IRAM& 12.0 \\
     &3$_{1,2}$--2$_{2,1}$ & 225.897 & 167.6 & $1.31 \times 10^{-5}$ & IRAM& 13.0 \\
     &4$_{2,2}$--4$_{2,3}$ & 143.727 & 319.2 & $2.80 \times 10^{-6}$ & IRAM& 17.0 \\
     &5$_{2,3}$--4$_{3,2}$ & 255.050 & 437.2 & $1.78 \times 10^{-5}$ & IRAM& 12.0 \\            
 \tableline
p-H$_2^{18}$O &1$_{1,1}$--0$_{0,0}$ & 1101.698 & 53.9 & $1.79 \times 10^{-2}$ & HIFI & 19.2\\
 \tableline
o-H$_2^{18}$O &2$_{1,2}$--1$_{0,1}$ & 1655.868 & 114.0 & $1.26 \times 10^{-2}$ & HIFI & 12.8\\
\tableline
\end{tabular}
\tablecomments{
Spectroscopic data from the JPL spectral line catalog \citep{Pickett98},
http://www.spec.jpl.nasa.gov.\\
CSO data: this work; IRAM data from \citet{Jacq90};
Herschel/HIFI data from \citet{Flagey13}.
}
\end{center}
\end{table}


\subsection{Source description}

The 1$_{0,1}$--0$_{0,0}$ ground-state rotational transition of
HDO at 465 GHz and the 1$_{1,0}$--1$_{0,1}$ transition at
509 GHz were observed toward three high-mass star forming regions
are listed in Table 2. All the sources are characterized by strong
millimeter continuum and mid-infrared emission, characteristic of the
early stage of high-mass star formation.

G34.26+0.15 is one of the best studied high-mass star-forming regions
in the Milky Way. Embedded within this molecular cloud is a hot core,
which exhibits strong H$_2$O maser emission and high
abundances of saturated molecules \citep{Macdonald96}; two unresolved
UCHII regions, labeled A and B; a more evolved H II
region with a cometary shape; and an extended, ring-like H II
region \citep{Reid85}. Based on narrow-band mid-infrared imaging of
the complex, \citet{Campb00}~concluded that the same star is
responsible for ionization of the cometary H II component (C)
and heating the dust, but is not interacting with the hot core seen in
the molecular emission. At a 12\arcsec\ resolution, \citet{Hunter98}
also found the peak of the 350 $\mu$m emission to coincide with the
component C of the UCHII region.

The radio continuum emission of W51 shows three separate components:
W51~IRS1, W51~IRS2, and W51~Main. W51~Main is defined by a group of OH
and H$_2$O masers near several UCHII regions. The
continuum emission from ultracompact H II regions was resolved
into compact components labeled W51e$_{1}$ to W51e$_{8}$
\citep{Zhang97}. Of these, e$_{1}$ and e$_{2}$ are the brightest in
high-resolution continuum maps.

The star forming region W49N contains at least a dozen UCHII
regions powered by OB-type stars arranged in a ring 2~pc in diameter
\citep{Pree00}. Evidence that star formation is still in progress
within W49N comes from strong H$_2$O maser emission and strong
millimeter continuum emission attributed to dust condensations
\citep{Sievers91}.


\begin{table}
\caption{Source sample.}
\label{}
\begin{tabular}{l ccc}
\tableline\tableline
Source & $\alpha$(J2000)& $\delta$(J2000) & $D$ \\
            &          &            & (kpc) \\
\tableline
G34.26+0.15 & 18 53 18.6 & +01 14 57.7 & 3.8\tablenotemark{a} \\
W51e$_{1}$/e$_{2}$ & 19 23 43.9 & +14 30 25.9 & 5.4\tablenotemark{b}\\
W49N & 19 10 13.2 & +09 06 12.0 & 11.4\tablenotemark{c}\\
\tableline
\end{tabular}
\tablenotetext{a}{\citet{Fish03}.}
\tablenotetext{b}{\citet{Sato10}.}
\tablenotetext{c}{\citet{Gwinn92}.}
\end{table}

\subsection{Observations}

Observations of the 465~GHz and 509~GHz HDO transitions presented here
were carried out in 2012 June--August, using the 10.4 m Leighton
Telescope of the Caltech Submillimeter Observatory (CSO) on Mauna Kea,
Hawaii. We used the new wideband 460 GHz facility SIS receiver and the
FFTS backend that covers the full 4 GHz intermediate frequency (IF)
range with a 270~kHz channel spacing. 
Pointing of the telescope was checked by performing five-point continuum scans of planets and strong dust continuum sources. The spectra were obtained in antenna temperature units $T^{*}_{A}$, and then converted to the main beam brightness temperature,  $T_{\mathrm{mb}}$, via the relation  $T^{*}_{A}$ = $\eta_{\mathrm{mb}}$$T_{\mathrm{mb}}$, $\eta_{\mathrm{mb}}$ is the CSO main-beam efficiency, which is found to be $\sim 37\%$ at 460~GHz from total power observations of planets. The absolute calibration uncertainty of the individual measurements is $\sim$20\%. \par
In addition to the new CSO data, we included in our analysis previously published observations of higher-energy transitions \citep{Jacq90} toward our target sources. We used the reduced HIFI data of the H$_2^{18}$O  transitions at 1101.698 and 1655.868 GHz. The HDO and H$_2^{18}$O lines parameters are listed in Table 3. The H$_2^{18}$O ground state transitions have been previously presented by \citet{Flagey13}.\par  
The data processing was done using the IRAM GILDAS software package \citep{Pety05}. We measured the line parameters: central velocity $V_{LSR}$, the full-width at half maximum (FWHM) $\Delta{V}$, peak intensity $T^{peak}$, by fitting a single gaussian profile to the data (in $T_{\mathrm{mb}}$ units). The integrated line intensity is equal to $W_{i}$ = $\int{T_i}dV$, where $i$ either refers to observations (obs) or models (mod).\par
We also used Herschel/HIFI data at 893 GHz (Vastel et al. in preparation) obtained by the PRISMAS guaranteed time key program, 
and the SCUBA data at 353 GHz that provide an accurate determination of the source continuum flux.\\ 
The PRISMAS continuum observations were obtained in the double beam switching mode. At 800 GHz, the HIFI 
 beam size is 26.5 $\arcsec$ and the instrument gain is 469 Jy/K~\citep{Roelfsema12}.


\begin{table}
\caption{HDO and H$_2^{18}$O lines parameters.}
\label{}
\begin{tabular}{l c c c c c c c }
\tableline \tableline
Source &Species & Frequency & $V_\mathrm{LSR}$ & $\triangle V$& $T^{peak}_{obs}$ & $W_{obs}$ & $W_{mod}$ \\ 
       &        &  (GHz)    &  (km s$^{-1}$)    &  (km s$^{-1}$) &  (K)   &  (K km s$^{-1}$) & (K km s$^{-1}$)\\
\tableline
G34.26 & HDO    &           &                  &               &        &                 & \\
\hline     
       &        & 143\tablenotemark{a} & 57.7 (0.2) & 6.9 (0.4) &  0.4 & 2.6 (0.2)& 1.7 \\
       &        & 225\tablenotemark{a} & 57.4 (0.3) & 6.6 (0.4) & 1.2 & 8.4(0.7) & 11.4\\
       &        & 241\tablenotemark{a} & 57.7 (0.3) & 7.0 (0.6) & 1.8 & 13.2(1.0) & 12.0\\
       &        & 255\tablenotemark{a} & 57.7 (0.9) & 8.8 (3.0) & 0.6 & 6.0 (0.6) & 3.4 \\
       &        & 465\tablenotemark{b} & 58.0 (0.1) & 6.4 (0.2)&  1.8 & 12.1(0.3) & 13.2\\
       &        & 509\tablenotemark{b} & 58.4 (0.2) & 6.0 (0.7)&  0.8 & 5.5 (0.8) & 10.4 \\
\hline
       &H$_2^{18}$O&  & & & & & \\

       &        & 1102\tablenotemark{c}& 61.1 (0.1) & 3.5 (0.2)& -0.9 & -3.4(0.2)& -2.6 \\
       &        & 1656\tablenotemark{c}& 61.1 (0.2) & 6.5(0.0)& -1.7 &  -11.7 (0.9)& -13.5\\
\hline\hline  
W51e$_{1}$/e$_{2}$& HDO   &  & & & & & \\
\hline
      &     & 225\tablenotemark{a} & 55.0 (0.4) & 9.2 (0.9) & 0.6  &  6.2(0.6)& 4.5 \\
      &     & 241\tablenotemark{a} & 55.6 (1.6) & 5.7 (1.6) & 0.8  & 4.7(1.1)& 5.5 \\
      &     & 255\tablenotemark{a} &  53.9 (2.0) & 15.6 (6.0) & 0.3  & 5.6(2.0)& 2.1\\
      &     & 465\tablenotemark{b} & 57.1 (0.2) & 6.4 (0.4) & 1.3 & 8.7(0.4) & 10.7 \\
      &     & 509\tablenotemark{b} &  &   & $<$ 0.9\tablenotemark{d}  & $<$ 3.5\tablenotemark{e} & 4.1  \\
\hline
       &H$_2^{18}$O  &  & & & & & \\

       &    & 1102\tablenotemark{c}& 58.3 (0.1) & 6.1 (0.3)& -0.9 & -5.8(0.3)& -5.7 \\
       &    & 1656\tablenotemark{c}& 58.7 (0.1) & 8.3(0.3)& -2.1 &  -18.3 (0.6)& -20.0\\ 
\hline\hline
W49N &  HDO &  & & & & & \\
\hline
       &   & 465\tablenotemark{b} & 8.5(0.2) & 12.0 (0.5) & 0.8 & 10.8 (0.4) & 9.4\\
       &   & 509\tablenotemark{b} & 8.1 (0.7)& 12.9 (1.5) & 0.7 & 10.2 (1.2) & 8.9\\
\hline
&H$_2^{18}$O &  & & & & & \\

      &     & 1102\tablenotemark{c}& 10.3 (0.2) & 9.6 (0.4)& -0.7 & -7.5(0.3)& -7.7\\
      &     & 1656\tablenotemark{c}& 10.4 (0.1) & 12.0(0.1)& -2.3 &  -30.2 (0.3)& -26.4\\ 
\hline
\end{tabular}
\tablenotetext{a}{\citet{Jacq90}.}
\tablenotetext{b}{This work.}
\tablenotetext{c}{\citet{Flagey13}.}
\tablenotetext{d}{$3\times rms$ upper limit}
\tablenotetext{e}{$3\sigma(\mathrm{K\,kms^{-1}})=3\times rms\times \sqrt{2\times d\nu \times \triangle V}$
 with $rms$ (root mean square) in $\mathrm{K}$, $d\nu$, the channel width in $\mathrm{km\,s^{-1}}$ and $\triangle V$, the $FWHM$ in $\mathrm{km\,s^{-1}}$. We assumed $\triangle V$ = 6.4 $\mathrm{km\,s^{-1}}$, which is the 465\,GHz emission line width.}

\end{table}



\section{Determination of the HDO and H$_2$O Abundance}

\subsection{ Modeling}

The goal of this study is to determine the HDO fractional
abundance in three high-mass stars formation regions. To
reproduce the observed line intensities (Table 3) the static radiative
transfer code of \citet{Zmuidzinas95} is used. The model cloud is
divided into 200 radial shells, and the code uses a multilevel
accelerated lambda-iteration method \citep{Rybicki91}~to solve for the
HDO level populations and the line and continuum radiative transfer in
a self-consistent fashion. This radiative transfer program takes into
account the excitation of HDO molecules by collisions, line
radiation, and dust continuum radiation at the HDO line frequencies.
However, IR radiative pumping through HDO vibrationally excited levels
and the large-scale velocity field, characteristic of infall or
expansion, are not included. The HDO collisional rates used
in this study were recently computed by \citet{Faure12} and
\citet{Wiesenfeld11} for ortho-H$_2$ and para-H$_2$ in the temperature range
5--300~K and for all rotational transitions with an upper energies
less than 444~K.
 In the modelling, we assumed a constant ortho-to-para ratio (OPR)   
 of H$_2$~equal to 3. The ortho and para H$_2$O collisional rates with ortho and para H$_2$
were taken from the LAMDA data base \citep{Schroier05, Daniel11}. These rates 
were calculated for temperatures in the range from 5 to 1500~K including 
energy levels up to 2000~K above ground. The same collisional rates are used for the H$_2^{18}$O isotopologue.
The limitation imposed by the radiative transfer code is to use  a single collisional 
partner in the calculations. We assume that all hydrogen is in the ortho state. \par

We carried out model calculations from the inner core radius, $r_{min}$,
to the outer radius, $r_{max}$ (with ${r_{max}}/{r_{min}} \sim 100$ for all
sources; \citealt{Hatchell03}). The distance of the edge of the core
from the star is set at $r_{min}\sim2.0\times 10^{16}$~cm or
$\sim1000$~AU, with no dust emission seen at the smaller radii. The
lack of submillimeter emission in the core centers could be due to
optical depth effects or a central cavity \citep{Tak00}.
We adopted a dust-to-gas ratio of 1:100 and a power-law
H$_2$ density distribution of the form:
\begin{eqnarray}
 n(r) =  n_0\bigg(\frac{r}{r_{min}}\bigg)^{-1.5} \qquad
\end{eqnarray}
\noindent where $n_0$ is the H$_2$ density at the reference
radius ($r_{min}=1000$~au). The power-law index was set to 1.5
in accordance with the static infall theory in the inner part of the object
\citep{Shu77, Tak00, Beuther02, Marseille10}. We assumed that the gas and
dust radial temperature profiles follow a power law \citep{Viti99}:

\begin{equation}
T(r) = T_0\bigg(\frac{r}{r_{min}}\bigg)^{-0.5}
\end{equation}

\noindent where $T_0$ represents the maximum temperature of dust
grains. We assumed that, at densities found in the hot cores, the gas
temperature is equal to the dust temperature.
\subsection{Dust emissivity index $\beta$}
When molecules deplete inside prestellar cores, dust emission may
represent the best tracer of the gas density distribution just prior
to the onset of gravitational collapse. The dust continuum optical
depth is described by a power-law frequency dependence, $\tau \propto
\nu_{}^{\beta}$, and to fit the observed spectral energy distribution,
the knowledge of the grain emissivity spectral index, $\beta$, is
required. The dust emissivity index depends on the dust grain
composition, size, and temperature \citep{Hild83, Goldsmith97}. Details
of the dust-modelling process can be found in the reviews by
\citet{Draine03}. Observationally, there have been
many attempts at determining and explaining $\beta$. Typical values of
$\beta$ range between 1 and 2, with further support for $\beta$ =
1.5--2.0 coming from observations: \citet{Wright92}, \citet{Minier05}
and \citet{Gordon10}.\,Planck Collaboration XIV (2013) used
  \textit{Planck} HFI data with ancillary radio data to study the
  emissivity index. They computed a median value of far infrared
  spectral index $\beta_{FIR}$ = $1.88~\pm~0.08$ at the high frequency
  \textit{Planck} channels ($\nu \geq 353$~GHz) and a median value of
  spectral index $\beta_{mm}$ = $1.6~\pm~0.06$ at millimeter
  wavelengths ($\nu < 353$~GHz). We can estimate the dust
  grain emissivity exponent from observations at two frequencies
  $\nu_1$ and $\nu_2$ \citep{Hill06}:
\begin{equation}
\beta= \frac{{\log\frac{F^{}_{\nu_2}}{F^{}_{\nu_1}}}+\log\frac{({\rm
      e}_{}^{h\nu_2/kT_{\mathrm{dust}}}-1)}{({\rm
      e}_{}^{h\nu_1/kT_{\mathrm{dust}}}-1)}}{{\log\frac{\nu_2}{\nu_1}}}-3 
\end{equation}
where $F^{}_\nu$ is the source flux density, $\nu$ the frequency of
the observations, and $T_{\mathrm{dust}}$ the dust temperature. In
this work, the dust grain emissivity index $\beta$ is determined for
G34.26 using the millimeter ($\lambda_1$ = 1.2~mm; $\nu_1$ = 250~GHz)
data obtained with SIMBA and submillimeter ($\lambda_2$ = 450~$\mu$m;
$\nu_2$ = 660~GHz) SCUBA data. We derive $\beta = 1.6 \pm 0.48$. The
uncertainty in $\beta$ is typically 30\% for the 20--50 K temperature
range \citep{Hill06}. The spectral index for the other sources is
taken from literature: \citet{Thompson90} for W49N and from
\citet{Gordon87} for W51. The HDO lines studied here are seen
  in emission, and the model intensities are not sensitive to the exact
  value of $\beta$, especially for $T > 30~\mathrm{K}$. That is why
  the dust grain emissivity index is fixed, and not a free parameter,
  in the fits.

The values of $\beta$ in our target sources, as well as the continuum
fluxes from SCUBA and Herschel/HIFI observations, are listed in
Table~4. These are used to constrain the density and temperature
distributions as input to the line modeling.

\subsection{Modeling procedure}
We approximated the radial variation of the HDO fractional
abundance, $X=n(HDO)/n(H_{2})$ as a step function
with an enhanced abundance $ X_{in}$ in the inner region where $T \geq
100~\mathrm{K}$, and a lower value $ X_{out}$ for the outer envelope
where $T < 100~\mathrm{K}$. Laboratory studies indicate that the
evaporation temperature lies in the 90--110~K range, depending on the ice
composition and structure \citep{Fraser05}. In this work the
sublimation temperature of water, $T=100$~K \citep{Fraser01}, is
applied as the jump temperature.
The model uses the following free parameters: $n_0$, $T_0$ and
$X_{in}$ (for $ T \geq 100~\mathrm{K}$) and $X_{out}$ (for $T <
100~\mathrm{K}$). We determine the continuum flux density per beam at
353 GHz (850~$\mu$m), 509 GHz (590~$\mu$m), and 893~GHz (336~$\mu$m)
from the model.
\begin{table}[tb]
\caption{Continuum flux densities and grain emissivity exponents}
\label{}
\begin{tabular}{l c c c c}
\tableline\tableline
 Source & $F_{353}$\tablenotemark{$\star$}\tablenotemark{a} & $F_{509}$\tablenotemark{b} &
 $F_{893}$\tablenotemark{b} & $\beta$ \\
        &  (Jy/beam) & (Jy/beam) & (Jy/beam)    \\

\tableline 
  G34.26 + 0.15 & 56.1 & 310 & 1320 & 1.6\tablenotemark{c}\\
  W51e$_{1}$/e$_{2}$ &  & 400 & 1490 & 1.7\tablenotemark{d}\\
  W49N &  & 320 & 1450  & 1.8\tablenotemark{e}\\
\tableline
\end{tabular}
\tablenotetext{\star}{$F^{}_\nu$ the source flux density, $\nu$ the frequency of the observations in GHz}
\tablenotetext{a}{SCUBA; the accuracy is about 6$\%$ \citep{Thompson06}.}
\tablenotetext{b}{Herschel; the accuracy is about 10$\%$. }
\tablenotetext{c}{SIMBA and SCUBA observations \citep{Hill06}.}
\tablenotetext{d}{Ward-Thompson \& Robson (1990).}
\tablenotetext{e}{Gordon et al. (1987).}

\end{table}

 Finally, the synthetic spectra and continuum emission are convolved to the
 appropriate telescope beam size for comparison with the
 observations. We minimize the `figure of merit' (FOM) with the method of 
 \citet{Jacq90}\,to find the best model of the source.
  The FOM is computed
from the observed and modeled spectra and continuum fluxes according
to the following formula:
\begin{equation}
 FOM
 = FOM_1 + FOM_2 =\sum_{n}\frac{(T_{\mathrm{obs}}-T_{\mathrm{mod}})^2}{(T_{\mathrm{obs}})^2}+\sum_{m}\frac{(F_{\mathrm{obs}}-F_{\mathrm{mod}})^2}{(F_{\mathrm{obs}})^2} 
\end{equation}
for a set of $n$ spectral lines and $m$ continuum flux densities.
The inner and outer HDO abundances are constrained by the
spectral line data ($FOM_1$), whereas model parameters describing the density and temperature 
distribution are constrained primarily by the continuum SED ($FOM_2$).
 To determine the uncertainty of $X_{in}$ and $X_{out}$, we performed $\chi^{2}$ analysis. In analogy to Lampton method \citep{Lampton76} we define $S$ $\equiv$ $FOM_1/\sigma^2$ $=$ $\sum_{n}\frac{(T_{\mathrm{obs}}-T_{\mathrm{mod}})^2}{(\sigma T_{\mathrm{obs}})^2}$, where $n$ being the number of spectral lines. The $\sigma$ within analysis $\chi^{2}$ includes a calibrations uncertainty of $20\%$ for all individual measurments. The difference, $\Delta{S}$ $\equiv$ $S_{true}-S_{min}$ is distributed as $\chi^{2}$ with $p$ degrees of freedom (here: $p$=2; $X_{in}$ and $X_{out}$). By $\Delta{S}$ $\sim$ $\chi^{2}_p$ ~($"\sim"$ - "is distributed as") we mean for any number L probability: $Prob~(\Delta{S}>L)$ = $Prob~(\chi^{2}_p>L)$ . With the limiting contour value $S_L$ defined as $S_{min}+ L $, $Prob~(\Delta{S}>S_L - S_{min})$ = $Prob~(S_{true}> S_L)$ = $Prob~(\chi^{2}_p>S_L - S_{min})$.  $Prob~(S_{true}> S_L)$ is the probality $\alpha$ of the contour failing to enclose the true value, hence $\alpha$ =  $Prob~(\chi^{2}_p>S_L - S_{min})$. The $\alpha$-point of $\chi^2$ distribution is defined by $\alpha$ $\equiv$ $Prob~[\chi^{2}_p > \chi^{2}_p(\alpha)]$. Significance $\alpha$ is $S_L$ = $S_{min} + \chi^{2}_p(\alpha)$. In this expression, $\chi^{2}_p(\alpha)$ is tabulated value of $\chi^2$ distribtion for $p$ degrees of freedom and significance $\alpha$. Equivalently, any one observation's contour has a confidence $C = 1 - \alpha$ of enclosing the true parameter vector. The required contours for significance are 1$\sigma$, 2$\sigma$, and 3$\sigma$, which respectively represent a confidence of 68.3$\%$, 95.4$\%$, and 99.7 $\%$ of enclosing the true value of $X_{in}$ and $X_{out}$. The contours correspond to $S_L = S_{min} + 2.17$, $S_L = S_{min} + 6.17$, and $S_L = S_{min} + 11.8$.\\
 We used models with
the same physical parameters as those used for HDO in the analysis of the 
H$_2^{18}$O data. The two H$_2^{18}$O fundamental transitions were modeled independently and
the resulting ortho/para ratio of water is consistent with the high  temperature 
value of given the modeling uncertainties .

\subsection{Results}

\subsubsection{Origin of The Lines}
Figure 1 shows fractional population of the HDO levels of the relevant
features of our dataset calculated in our model. The high-energy transitions are sensitive to changes in $X_{in}$.
Indeed the bulk of emission in the high energy transitions is produced in the inner 
\begin{figure}
\centering
\includegraphics[width=0.7\columnwidth]{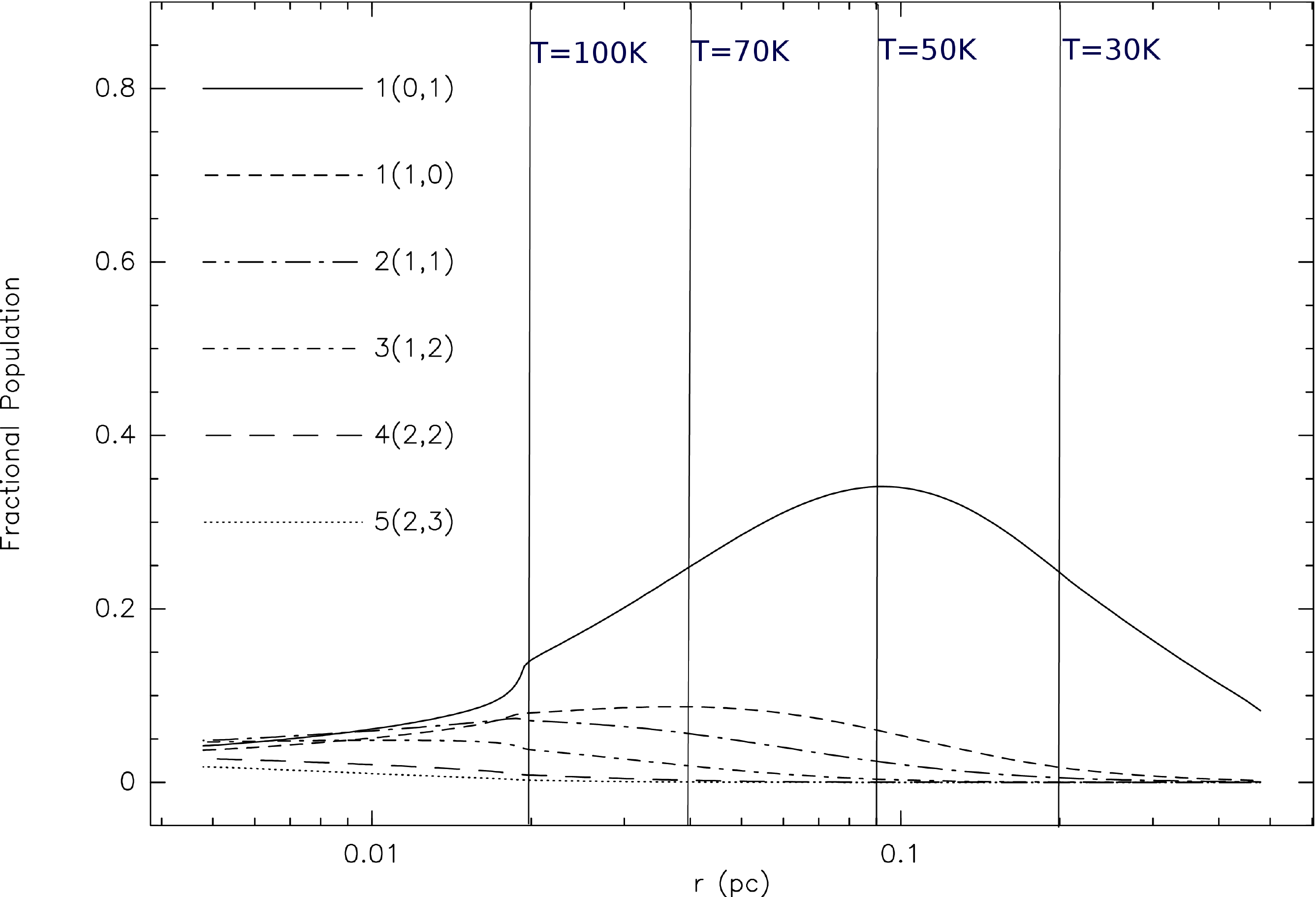}
\caption{Fractional population of the HDO energy levels as a function of the radius of the cloud.}
\label{}
\end{figure}
 hot-core region where $T > 100~\mathrm{K}$. In the G34.26
source, this region has a radius equal to 1.0\arcsec~which corresponds to
0.02~pc. This is in agreement with the interferometric observations 
of the HDO lines at 225 and 241 GHz by \citet{Liu13}.
That is the reason why the high-energy transitions are
sensitive to changes in $X_{in}$. On the other hand, the 465 and
509 GHz HDO transitions are sensitive primarily to
$X_{out}$. The 509 GHz line arises predominantly in the region between the warm envelope
and the cold region ($T \sim$ 50--100~$\mathrm{K}$), whereas the 465 GHz transition is produced in the 
cold envelope ($T < 50~\mathrm{K}$). The ground-state rotational transition of HDO at 465 GHz is consequently 
a very good probe of the abundance in the cold outer envelope, which is consistent with the results of \citet{Parise05}
for the solar-type protostar IRAS 16293-2422. The 509 GHz transition provides particurlarly good constraints on the HDO abundance
profile in the transition region between the hot core and the envelope, and should be included in future, more advanced models of HDO in high-mass star-forming regions. The model reproduces the observed intensities of different transitions in our target sources, with the exception of the 509 GHz line in G34.26.
Although the signal-to-noise ratio of the observed 509~GHz
 spectrum is limited, it is clear that the best-fit model does not
 reproduce this line profile.
The 509 GHz transition is formed in the part of the cloud where, within the scenario proposed by \citet{Rolffs10},
various feedback: thermal, radiative, or turbulent mechanisms are expected during the process of massive stars 
formation.  In particular, it should be noted that, with the inclusion of velocity fields in their model, \citet{Coutens14} succesfully reproduced the 509~GHz line observed with \textit{Herschel}/HIFI toward this same source. 
 Other possibility is an accretion disk that is fed by the infalling envelope. This is also supported by observations made by \citet{Keto87} of G34.26.
 The size of this possible disk is about 9000~au (0.05~pc) \citep{Garay90, Hajigholi16} and agrees well with the place where the 509 GHz line arises (see Figure 1). This model of the G34.26 source \citep{Hajigholi16} neither confirm or refute the presence of an expansion in the inner parts of the envelope \citep{Coutens14}. We concluded that the geometry and physical structure of our model is too simplistic, and that is why we could not to reproduce the 509 GHz line.

\subsubsection{Target sources}
%
\textit{G34.26+0.15}:\,Observed  spectra (black line) and gaussian fit (blue line)~of the 465 and 509 GHz HDO transitions toward G34.26+0.15 along with the best-fit model (red line) are shown in Figure 2. 

\begin{figure}[h!]
 \begin{center}
  \begin{minipage}{0.48\textwidth}
\centering
    \includegraphics[width=\textwidth]{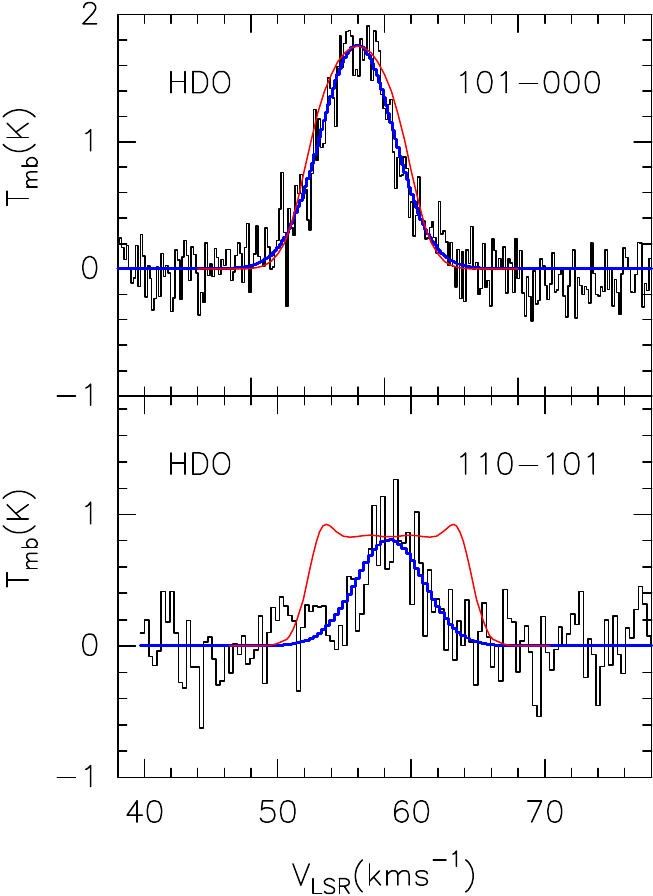}
      \end{minipage}
\quad
\vspace{0.5cm}
\begin{minipage}{0.48\textwidth}
\centering
    \includegraphics[width=\textwidth]{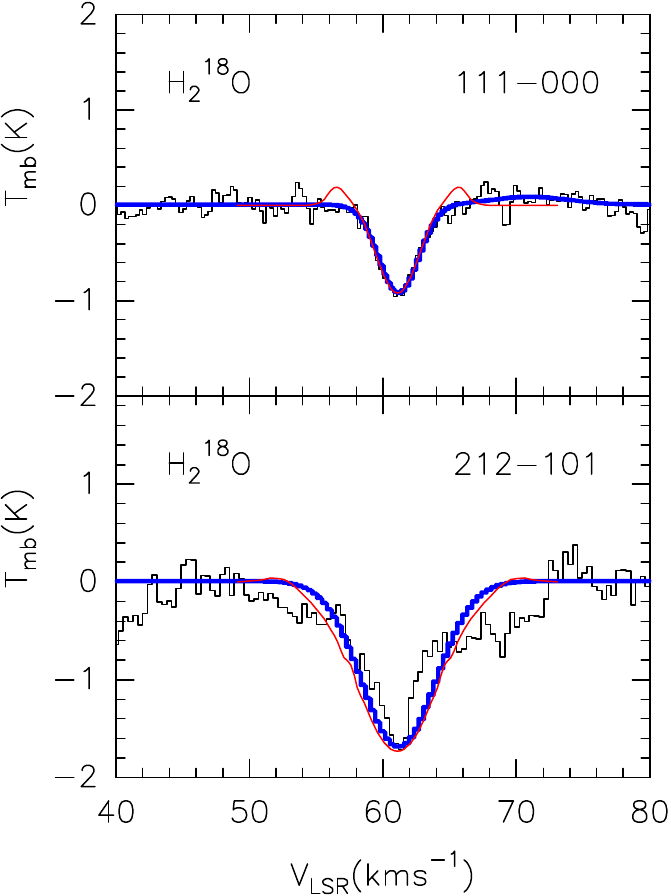}
 \end{minipage}
 \label{}
    \caption{Observed spectra of the 465 and 509 GHz HDO lines and 1102 and 1656 GHz H$_2^{18}$O lines towards G34.26+0.15.
             The Gaussian fits are shown in blue, while the best-fit model in red.}
\end{center}
 \end{figure}

 Model results are presented in Table 5. We obtain the
best--fit model for: $T_{0}=200$~K, $n_{0} = 1.0\times
10^{8}\,cm^{-3}$,\, $X_{in}=3.7\times 10^{-8}$ and $X_{out}=7.8\times
10^{-11}$. We calculated continuum flux densities at 353, 509,
and 893 GHz. The resulting uncertainties of $X_{in}$ and $X_{out}$ are shown in Figure 3 
and listed in Table 8.\\
\begin{figure}[ht]
\begin{center}
\includegraphics[width=0.7\columnwidth]{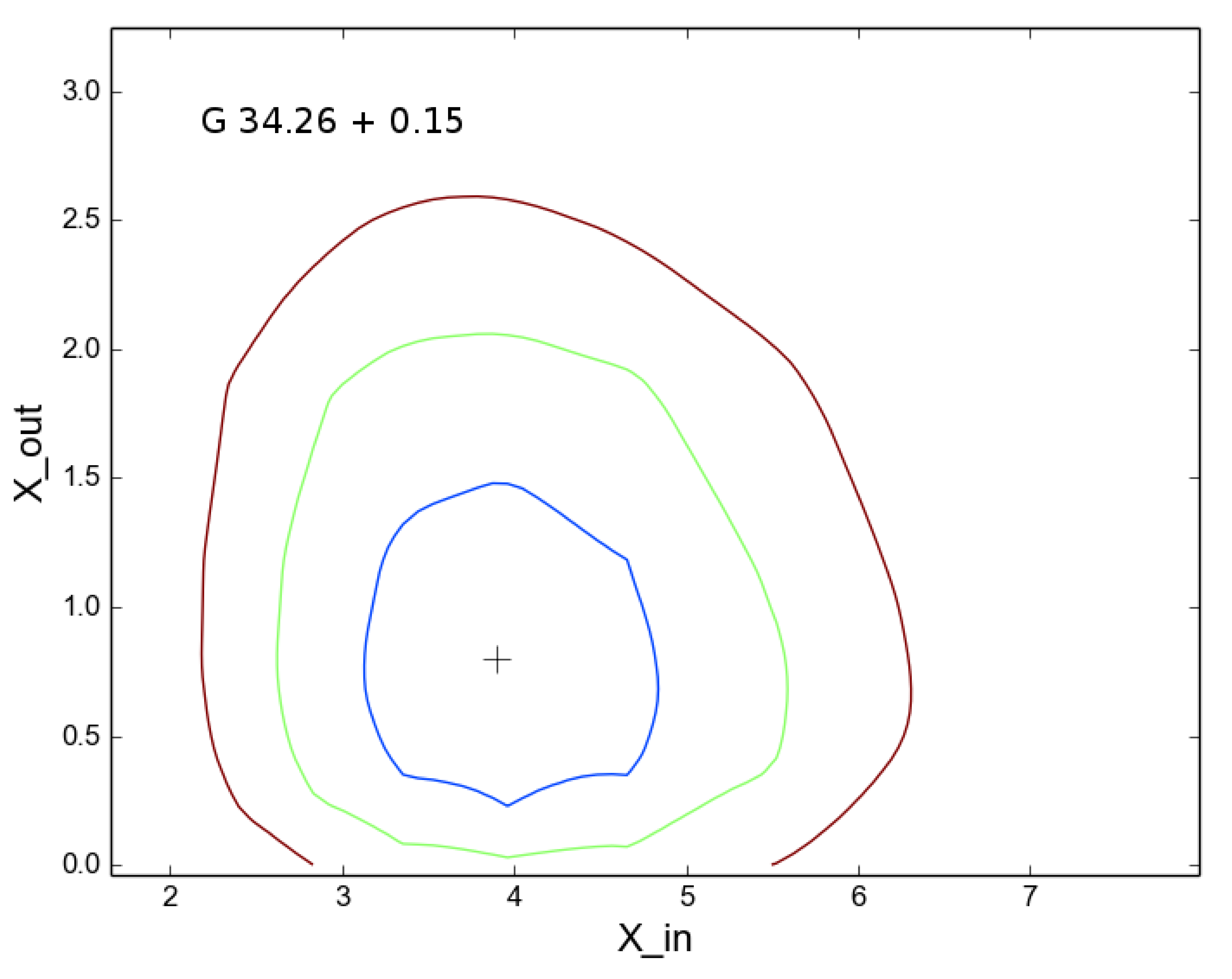}
\caption{$X_{in}$ and $X_{out}$ HDO abundance contours at 1$\sigma$, 2$\sigma$, 3$\sigma$ for $\chi^2$.
 The best-fit model is represented by the symbol $"+"$ ($X_{in}$ = a$\times 10^{-8}$, $X_{out}$ = b$\times 10^{-10}$).}
\label{}
\end{center}
\end{figure}
 Observed and modeled spectra of the para-H$_2^{18}$O line at 1102 GHz 
and the ortho-H$_2^{18}$O line at 1656 GHz are shown in Figure 2.  The derived OPR in G34.26 is 1.9.
The total (ortho+para) H$_2^{18}$O  abundance in the envelope $X_{out}$, is 4.9$\times
10^{-11}$. The recommended isotopic abundance ratio between $^{16}$O and $^{18}$O is 500 \citep{Lodders03}.
 Using this value, the H$_2$O outer abundance is 2.5$\times 10^{-8}$ and the outer HDO/H$_2$O ratio is 3.1$\times 10^{-3}$
in the  envelope. Considering the results with a 20$\%$ calibration uncertaint, the outer abundance ratio is (2.5 - 3.7)$\times 10^{-3}$.

\begin{table}
\caption{Line intensities and continuum fluxes for the best fit model of G34.26+0.15.}
\label{}
\begin{tabular}{l c c c c r}
  \tableline\tableline
  HDO & Freq  & $T_\mathrm{obs}$ &$T_\mathrm{mod}$ & $FOM_1$ \\
  transitions    &  (GHz)   &  (K) &    (K)   &    & \\      
  \hline
  1$_{0,1}$--0$_{0,0}$ & 464.925\tablenotemark{a} & 1.8 & 1.8 & 0.00 \\
  1$_{1,0}$--1$_{0,1}$ & 509.292\tablenotemark{a} & 0.8 & 0.8 & 0.00 \\
  2$_{1,1}$--2$_{1,2}$ & 241.562\tablenotemark{b} & 1.8 & 1.6 & 0.01 \\
  3$_{1,2}$--2$_{2,1}$ & 225.897\tablenotemark{b} & 1.2 & 1.4 & 0.03 \\
  4$_{2,2}$--4$_{2,3}$ & 143.727\tablenotemark{b} & 0.4 & 0.5 & 0.06 \\
  5$_{2,3}$--4$_{3,2}$ & 255.050\tablenotemark{b} & 0.6 & 0.4 & 0.11 \\
  \hline\hline
 Flux (Jy/beam) &  $F_{353}$\tablenotemark{c} & $F_{353_\mathrm{mod}}$ &  & $FOM_2$ \\
     
       &   56.1      &      56.4        &     & 0.00\\
\hline
       &  $F_{509}$\tablenotemark{d} & $F_{509_\mathrm{mod}}$ &  & \\
       
       &   310      &      376       &     & 0.05\\
\hline
       &  $F_{893}$\tablenotemark{d}  & $F_{893_\mathrm{mod}}$&   &  \\
      
       &   1320     &     1525        &     &  0.02\\
       &            &                 &     &  $FOM$ = 0.28 \\
 \tableline
\end{tabular}
\tablenotetext{a}{This work.}
\tablenotetext{b}{\citet{Jacq90}.}
\tablenotetext{c}{CSO}
\tablenotetext{d}{Herschel}

\end{table}



%
\begin{figure}
 \begin{center}
  \begin{minipage}{0.48\textwidth}
\centering
    \includegraphics[width=\textwidth]{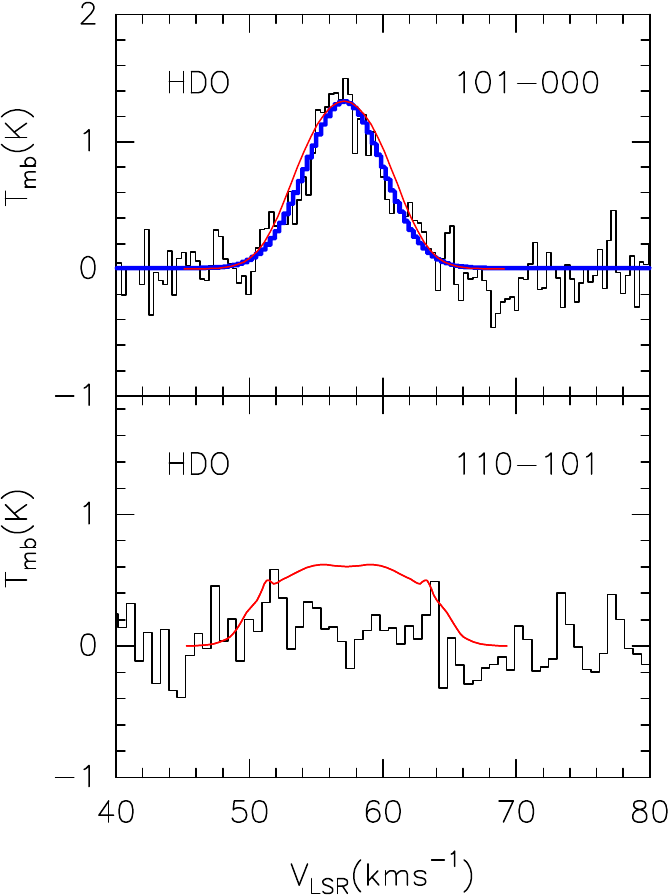}
      \end{minipage}
\quad
\vspace{0.5cm}
\begin{minipage}{0.48\textwidth}
\centering
    \includegraphics[width=\textwidth]{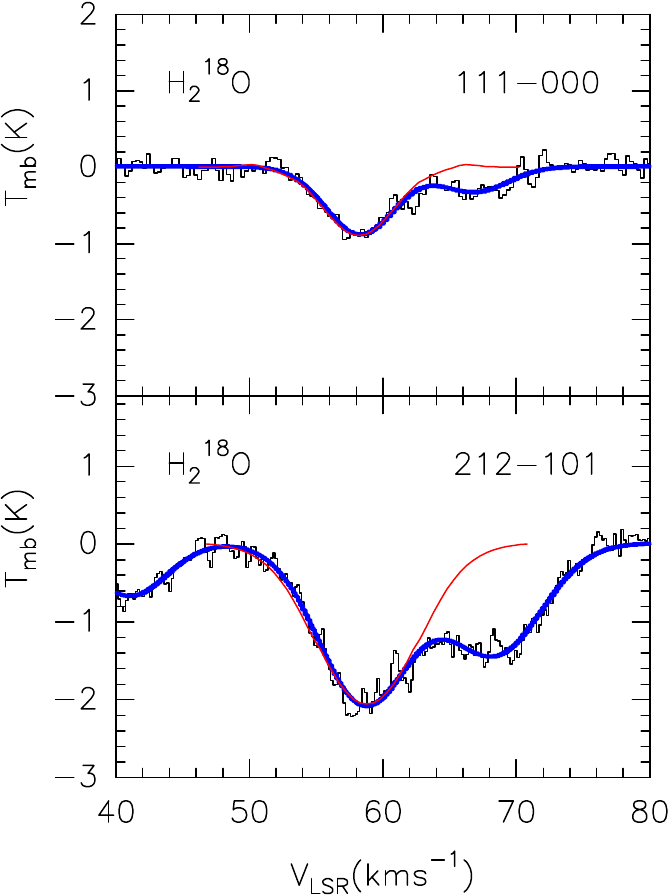}
 \end{minipage}
 \label{}
    \caption{Observed spectra of the 465 and 509 GHz HDO lines and 1102 and 1656 GHz H$_2^{18}$O
    lines toward W51e$_{1}$/e$_{2}$.The Gaussian fits are shown in blue, while the best-fit model in red.}
\end{center}
 \end{figure}

\textit{W51e$_{1}$/e$_{2}$}:\,Observed  spectra and gaussian fit of the 465  and 509 GHz HDO transitions toward W51e$_{1}$/e$_{2}$ along with the best-fit model, are shown in Figure 4 by black, blue and red lines, respectively.
 Model results for W51 are presented in
Table 6. We obtain the best fit for: $T_{0}=230~\mathrm{K}$,\,$n_{0} =
1.8\times 10^{ 8}\,cm^{-3}$,\, $X_{in}=1.7\times 10^{-8}$ and
$X_{out}=7.0\times 10^{-11}$. The resulting uncertainties of $X_{in}$ and $X_{out}$ are shown in Figure 5 
and listed in Table 8.\\Model flux densities per beam at 509 GHz
and 893 GHz for W51 are also listed in Table 6.\\
\begin{figure}[ht]
\begin{center}
\includegraphics[width=0.7\columnwidth]{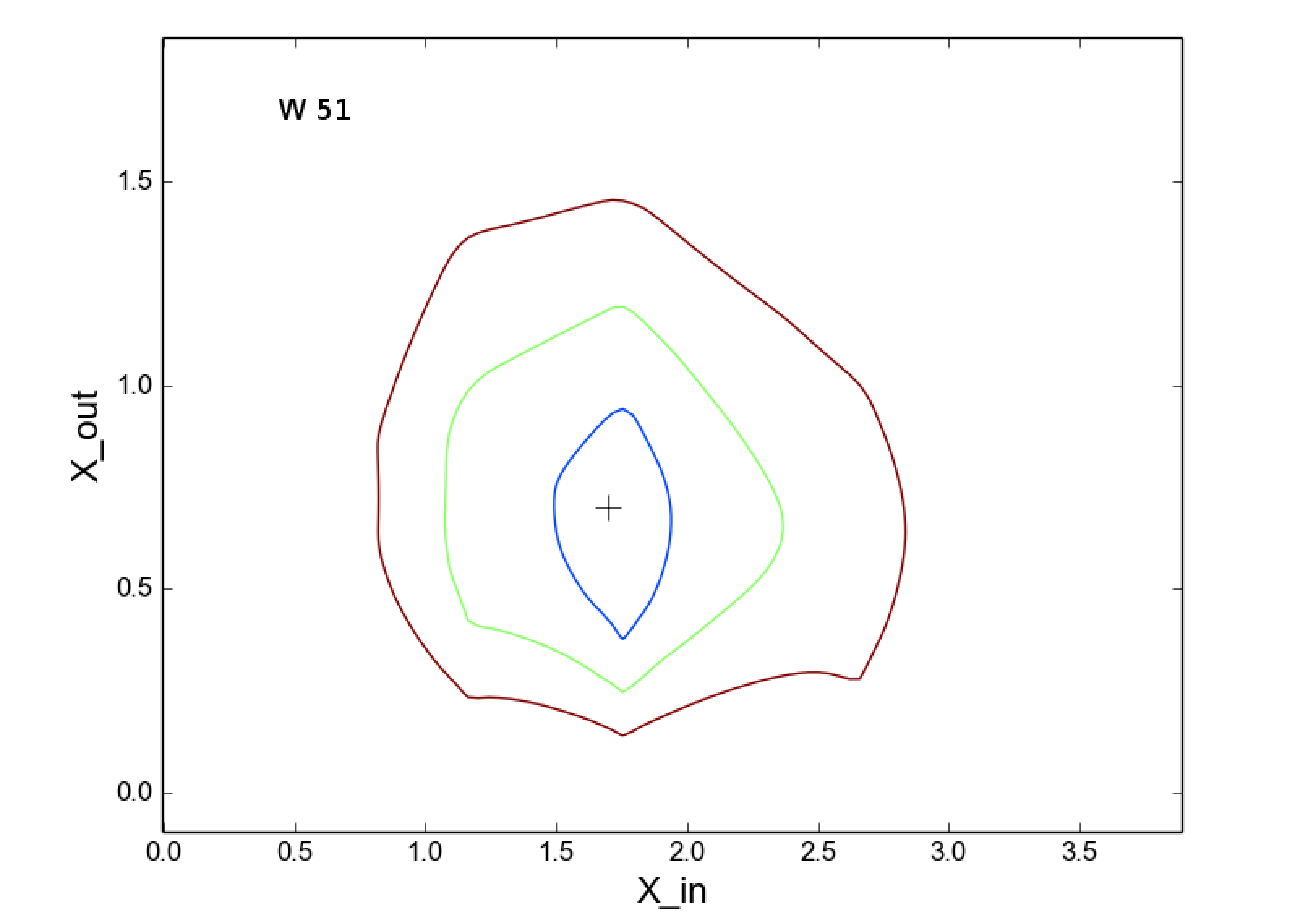}
\caption{$X_{in}$ and $X_{out}$ HDO abundance contours at 1$\sigma$, 2$\sigma$, 3$\sigma$ for $\chi^2$.
The best-fit model is represented by the symbol $"+"$ ($X_{in}$ = a$\times 10^{-8}$, $X_{out}$ = b$\times 10^{-10}$).}
\label{}
\end{center}
\end{figure}

 Observed and modeled spectra of the para-H$_2^{18}$O line at 1102 GHz 
and the ortho-H$_2^{18}$O line at 1656 GHz are shown in Figure 4.
The total H$_2^{18}$O abundance (OPR = 2.9) in the envelope $X_{out}$, is $5.5\times
10^{-11}$. The H$_2$O outer abundance is 2.8$\times 10^{-8}$ and outer HDO/H$_2$O ratio is 2.5$\times 10^{-3}$.
Considering the results with the 20$\%$ calibration uncertainty the outer ratio is (2.0 - 3.0)$\times 10^{-3}$.

\begin{table}
\caption{Line intensities and continuum fluxes for the best fit model of W51e$_{1}$/e$_{2}$.}
\label{}
\begin{tabular}{l c c c c r}
\tableline\tableline
HDO & Freq  & $T_\mathrm{obs}$ & $T_\mathrm{mod}$ & $FOM_1$ \\
transitions    &  (GHz)   &  (K) &    (K)   &      \\      
\tableline
1$_{0,1}$--0$_{0,0}$ & 464.925\tablenotemark{a} & 1.3 & 1.3 & 0.00 \\
2$_{1,0}$--1$_{0,1}$ & 509.292\tablenotemark{a} & $<$0.9 & 0.6 &  \\
2$_{1,1}$--2$_{1,2}$ & 241.562\tablenotemark{b} & 0.8 & 0.8 & 0.00 \\
3$_{1,2}$--2$_{2,1}$ & 225.897\tablenotemark{b} & 0.6 & 0.67 & 0.01 \\
5$_{2,3}$--4$_{3,2}$ & 255.050\tablenotemark{b} & 0.3 & 0.2 & 0.11 \\
\hline\hline
 Flux (Jy/beam)  &  $F_{509}$\tablenotemark{c} & $F_{509_\mathrm{mod}}$ &  & $FOM_2$ \\
       
       &  400   &    350  &  &  0.02 \\
\hline
       &  $F_{893}$\tablenotemark{c}  & $F_{893_\mathrm{mod}}$&   &  \\
       
       &   1490     &     1509       &     &  0.00\\
       &            &                 &     &  $FOM$ = 0.14 \\
 \tableline
\end{tabular}
\tablenotetext{a}{This work.}
\tablenotetext{b}{\citet{Jacq90}.}
\tablenotetext{c}{Herschel}

\end{table}

%
\textit{W49N}:\,Observed  spectra and gaussian fit of the 465 and 509 GHz HDO transitions toward W49N along with the best-fit model are shown in Figure 6 by black, blue and red lines, respectively.
Model results for W49N are presented in Table 7.We obtain the best fit
for: $T_{0}=300~\mathrm{K}$,\,$n_{0} = 2.5\times 10^{8}\,cm^{-3}$,\,
$X_{in}=0.3\times 10^{-8}$ and $X_{out}=10 \times 10^{-11}$. The resulting uncertainties of $X_{in}$ and $X_{out}$ are shown in Figure 7 
and listed in Table 8.\\ As data
on the high excitation lines are missing for W49N, the inner abundance
$X_{in}$ is not as well constrained as in the other sources.\\

\begin{figure}[!ht]
 \begin{center}
  \begin{minipage}{0.48\textwidth}
\centering
    \includegraphics[width=\textwidth]{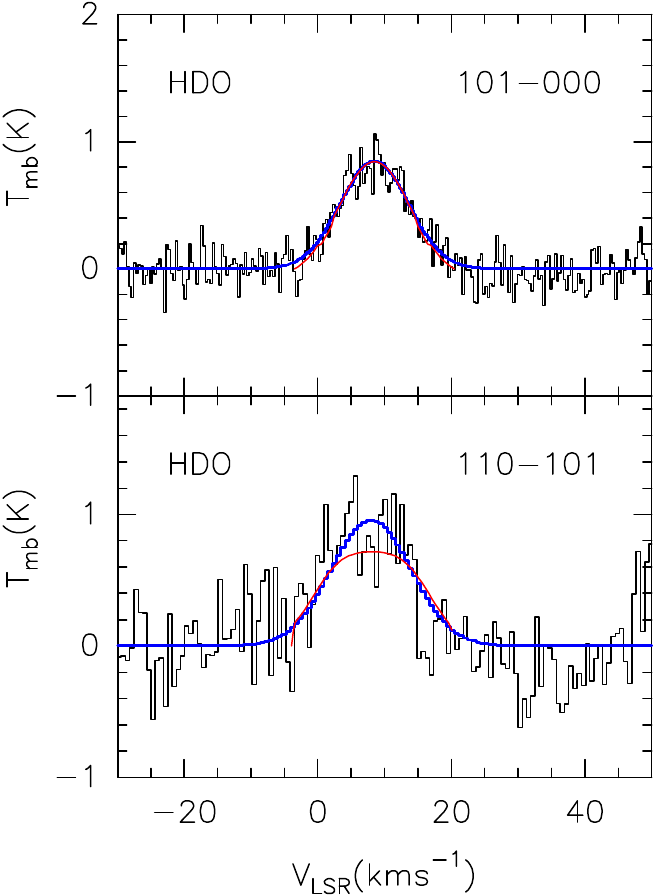}
      \end{minipage}
\quad
\vspace{0.5cm}
\begin{minipage}{0.48\textwidth}
\centering
    \includegraphics[width=\textwidth]{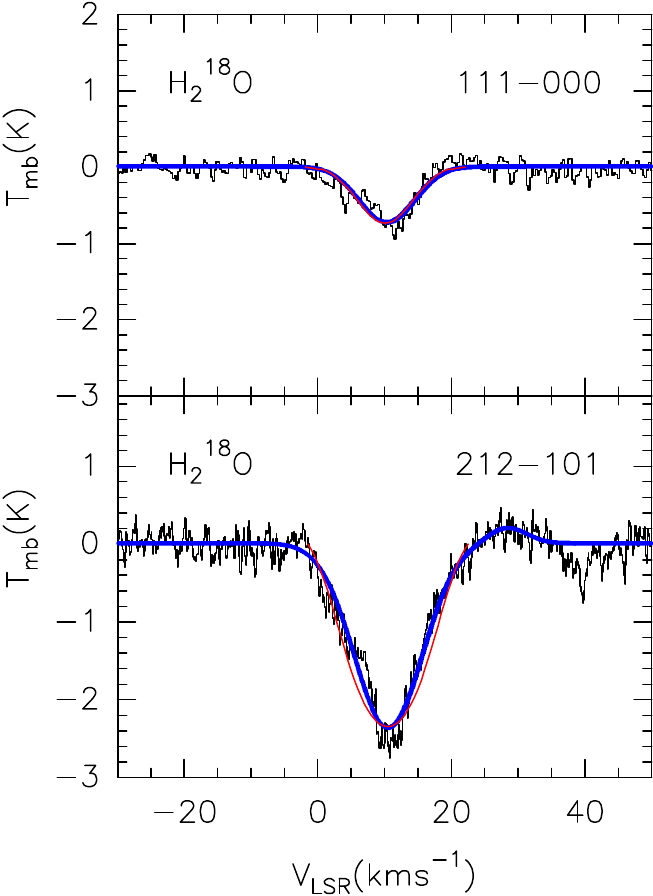}
 \end{minipage}
 \label{}
    \caption{Observed spectra of the 465 and 509 GHz HDO lines and 1102 and 1656 GHz H$_2^{18}$O lines toward W49N.
             The Gaussian fits are shown in blue, while the best-fit model in red.}
\end{center}
 \end{figure}


Observed and modeled spectra of the para-H$_2^{18}$O line at 1102 GHz 
and the ortho-H$_2^{18}$O line at 1656 GHz are shown in Figure 6.
The total H$_2^{18}$O (OPR = 3.1) abundance in the envelope $X_{out}$, is $1.1\times
10^{-10}$. The H$_2$O outer abundance is 5.5$\times 10^{-8}$ and the outer HDO/H$_2$O  ratio is 1.8$\times 10^{-3}$.
 Considering the results with the 20$\%$ calibration uncertainty, the outer abundance ratio is (1.4 - 2.2)$\times 10^{-3}$.

\begin{figure}
\begin{center}
\includegraphics[width=0.7\textwidth]{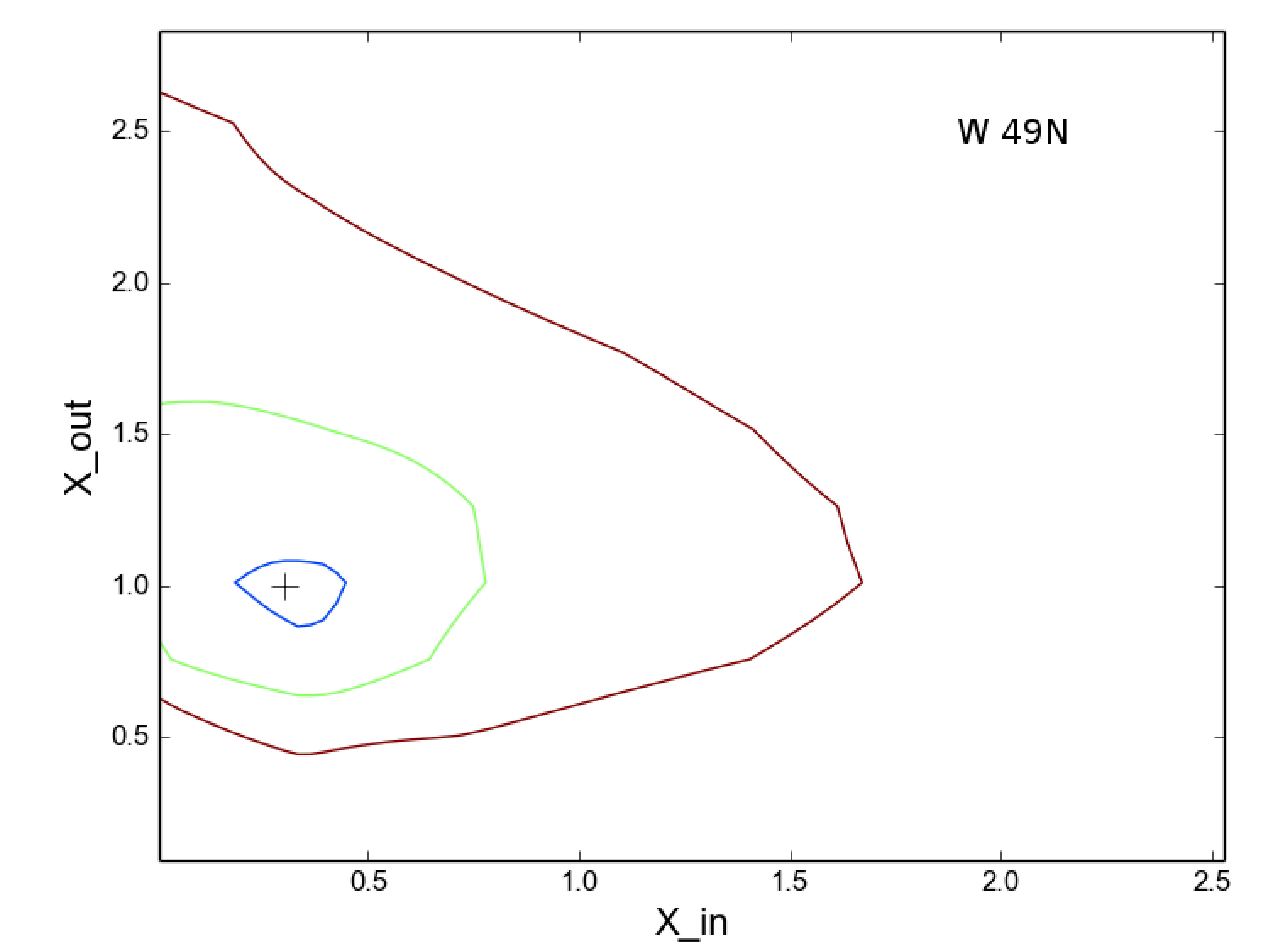}
\caption{$X_{in}$ and $X_{out}$ HDO abundance contours at 1$\sigma$, 2$\sigma$, 3$\sigma$ for $\chi^2$. The best-fit model is represented by the symbol $"+"$ ($X_{in}$ = a$\times 10^{-8}$, $X_{out}$ = b$\times 10^{-10}$).}
\label{}
\end{center}
\end{figure}



\begin{table*}
\caption{Line intensities and continuum fluxes for the best fit model of W49N.}
\label{}
\begin{tabular}{l c c c c r }
\tableline\tableline
HDO & Freq  & $T_\mathrm{obs}$ & $T_{\mathrm{mod}}$ & $FOM_1$ \\
transitions    &  (GHz)   &  (K) &    (K)   &      \\      
\hline
1$_{0,1}$--0$_{0,0}$ & 464.925\tablenotemark{a} & 0.8 & 0.8 & 0.00 \\
2$_{1,0}$--1$_{0,1}$ & 509.292\tablenotemark{a} & 0.7 & 0.6 & 0.02\\
\hline\hline
 Flux (Jy/beam) &  $F_{509}$\tablenotemark{b} & $F_{509_\mathrm{mod}}$ &  & $FOM_2$ \\
       
       &   320      &      310        &     & 0.00\\
\hline
       &  $F_{893}$\tablenotemark{b}  & $F_{893_\mathrm{mod}}$&   &  \\
       
       &   1450     &     1410        &     &  0.01\\
       &            &                 &     &  $FOM$ = 0.03 \\
\tableline
\end{tabular}
\tablenotetext{a}{This work.}
\tablenotetext{b}{Herschel}
\end{table*}

\section{Discussion}
\subsection{Comparison with Previous Studies}

Previous observations of the high-mass star-forming regions indicate an HDO abundance in the hot cores (T $>$ 100 K) but not in the outer, cooler envelopes (T $<$ 100K) (Jacq et al. 1990, Gensheimer et al. 1996, Pardo et al. 2001; see Table 9). We could find  variations in the derived HDO abundances in the hot cores between $1.5\times 10^{-9}$ and $2.0\times 10^{-7}$. Jacq et al. (1990) belived that, independent of all modeling, a value  much lower than  $3.0\times 10^{-8}$ for  HDO/H$_2$ in the hot core is very unlikely. For the first time, Comito et al. (2003, 2010) estimated the HDO abundance in both the inner and outer region of the high-mass source Sgr B2(M). These are, respectively: $3.5\times 10^{-9}$ (T$>$200K), $1.5\times 10^{-9}$ (100~K$<$T$<$200~K)  and  $2.5\times 10^{-11}$ (T$<$100K). The singly deuterated form of water has been also observed in the massive source AFGL 2591, with abundance varying
 from $1\times 10^{-7}$ in the hot core and $4\times 10^{-9}$ in the outer envelope \citep{Tak06}. \cite{Liu13} and \cite{Coutens14} determined the HDO abundance and HDO/H$_2$O ratio in the inner and outer region of G34.26 (see Table 9).
We derived  the HDO abundances of $X_{in}$ and $X_{out}$ in three high-mass star-forming regions: G34.26, W 49N, and W51. We found a difference between our $X_{in}(HDO)$ and $X_{out}(HDO)$ values for G34.26 and those reported by \citet{Coutens14}  and \citet{Liu13}, respectively. This is likely because the first authors used different model structures and a higher jump temperature, and  the second authors did not check the higher value  of $X_{out}(HDO)$ in their model.  The obtained HDO abundances 
of our target sources in the hot cores and the cooler envelopes are relatively consistent with the values found in the other high-mass star-forming regions \citep{Kulczak16}.
These results show that the HDO abundance is enriched in the inner regions of  high-mass protostars because of the sublimation of  the ice mantles, in the same way as for other studies low- and high-mass sources (e.g. NGC 1333 IRAS2A, IRAS 162923-2422, AFGL 2591, G34.26; Table 9). Observations of sites of high-mass star formation show in general the lower HDO abundances than observations of low-mass star forming cores. Possibly for high-mass protostars, the very cold and dense pre-collapse phase where CO freeze-out onto the grain mantles lasts only a short time, and the chemical reactions leading to the enhancement of deuterium abundance being strongly depressed when the temperature increases \citep{Caselli08}.

 \subsection{Variation of the HDO/H$_2$O Ratios with the Radius}

Based on observations of two H$^{18}_{2}$O fundamental transitions, we found that the H$_2$O abundances in our target sources are $(2.5 - 5.5)\times 10^{-8}$. Similar values were found  for the other high-mass protostars: $5\times 10^{-10}$ - $4\times 10^{-8}$ (\citealt{Marseille10};  \citealt{Herpin12}; \citealt{Tak10}; \citealt{Choi15}). The H$_2$O abundance in the cold envelope agrees fairly well with the model predictions for cold regions  where freeze-out takes place (\citealt{Ceccarelli96}; \citealt{Tak13}).\\  The water-deuterium fractionation in the inner and outer envelope of the high-mass star-forming region G34.26 was previously estimated by \citet{Liu13} and \citet{Coutens14}. We determined 
 the outer HDO/H$_2$O ratio in G34.26 to be $3.1\times 10^{-3}$, this value is relatively consistent with Coutens et al. (2014) (see Table 9).
 To estimate the inner HDO/H$_{2}$O ratio for the target sources, 
 we used an inner H$_{2}$O abundance value as high as $ ~ 10^{-4}$ from observations of other  high-mass star forming-regions 
  (\citealt{Boonman03}; \citealt{Snell00}; \citealt{Chavarria10}; \citealt{Herpin12}; \citealt{Tak13}). However a lower value of $\sim$ $10^{-6}$ was found in NGC 6334 I \citep{Emprechtinger13}. A possible explanation for the low water abundance in this source is a time-dependent effect; water molecules may not have enough time to fully desorp from the dust grain. Our derived  HDO/H$_2$O ratios are consequently not well-constrained.  We estimated that the inner HDO/H$_2$O  ratio is about $(1 - 4) \times 10^{-4}$  within the range found in other high-mass star-forming regions (\citealt{Jacq90}; \citealt{Gensheimer96}; \citealt{Emprechtinger13}; \citealt{Liu13}).   
 The HDO/H$_{2}$O  ratio varies between the inner and outer regions of high-mass protostars. The water deuterium fractionation decreases from the cold outer regions to the warm inner regions. The same trend is also present in low-mass protostars (Coutens et al. 2013 and 2014).
 The difference could be explained by the gradient
 of deuteration within interstellar ices. Only the external ice layers evaporate 
 in the cold envelope through non-thermal processes, whereas the inner part 
of ice mantles evaporates only in the hot core \citep{Taquet14}. 
The HDO/H$_2$O ratio in the bulk of ice mantle preserves the past physical and chemical conditions which materials experienced, 
while the HDO/H$_2$O ratio in active surface layers reflects local physical and chemical conditions (Furuya et al. 2015). The enrichment of deuterium in water ice should mostly occur in the latter prestellar core and/or protostellar phases, where interstellar UV radiation is heavily attenuated and CO is frozen out. Another possibility for the decrease of water deuterium fractionation toward the inner regions would be the additional water vapor formation at high temperatures ($T > 200 - 300~K$) thorough reactions: $O + H_{2} \longrightarrow OH + H$ and  $OH + H_{2} \longrightarrow H_{2}O + H$, which would decrease the HDO/H$_{2}$O ratios. However, it requires that a large amount of oxygen is in atomic form rather than in molecules in the high density inner regions.

\section{Summary}

\begin{table}
\caption{Best fit model parameters.}
\label{}
\begin{tabular}{l l l l c l c c }
\tableline\tableline
 Source  & $n_{0}(cm^{-3})$ & $T_{0}(\mathrm{K})$ & $X_{in}(HDO)$ & Range ($2\sigma$) & $X_{out}(HDO)$ &  Range ($2\sigma$)\\      
\hline
G34.26+0.15 &$1.0 \times 10^{8}$ & $200$ &$3.7 \times 10^{-8}$ & $2.7-5.6\times 10^{-8}$ &$7.8 \times 10^{-11}$ & $0.2-2.1\times 10^{-10}$  \\
W51e$_{1}$/e$_{2}$ & $1.8 \times 10^{8}$ & $230$ &  $1.7 \times 10^{-8}$ & $1.1-2.4\times 10^{-8}$ & $7.0 \times 10^{-11}$ & $0.3-1.1\times 10^{-10}$ &\\
 W49N&$2.5 \times 10^{8}$& $300$ & $0.3 \times 10^{-8}$ & $0.2-0.5\times 10^{-8}$\tablenotemark{$\star$} & $1.0 \times 10^{-10}$ & $0.8-1.6\times 10^{-10}$\\
 \tableline
 \end{tabular}
\tablenotetext{\star}{Range ($1\sigma$)}
 \end{table}

Using CSO observations of HDO low-excitation transitions, as well as previous observations of HDO high excitations 
 and H$^{18}_2$O low-excitation transitions from the literature, we determined the inner and outer HDO abundances, as well as  
the HDO/H$_2$O outer ratios toward three high-mass star-forming regions: G34.26 + 0.15, W51e$_{1}$/e$_{2}$, W49N. We derived
  HDO abundances of  $X_{in}$ = (0.3--3.7)\,$\times 10^{-8}$ (for $ T \geq 100~\mathrm{K}$) and $X_{out}$ = (7.0--10.0)\,$\times 10^{-11}$  (for $ T < 100~\mathrm{K}$),\, (see Table 8), and HDO/H$_2$O outer 
ratios of (1.8--3.1)\,$\times 10^{-3}$ (see Table 9). 
With this study, we showed that the 509 GHz transition can provide good constraints on the HDO abundance in the transition region between the hot core and colder envelope, and  that the 465 GHz HDO transition is a very good probe of the outer envelope of  massive protostars. These transitions could help for more advanced modeling of water in high-mass sources. The HDO/H$_2$O ratios were also found to be higher in the cold outer envelopes than in the hot cores, as already determined for two high mass sources.
What is important the model is very simple, easy to  implement, and not GPU-intensive,
and  provides a starting point for more sophisticated analysis.

\begin{table}
\scriptsize
\caption{Comparison of HDO abundance between different sources}
\label{}
\begin{tabular}{l c c c c c cr}
\tableline \tableline
Source  &  $X_{in}(HDO)$ (best fit) & $X_{out}(HDO)$ (best fit)& $(HDO/H_{2}O)_{in}$ & $(HDO/H_{2}O)_{out}$ & Ref.  \\ 
\tableline
\multicolumn{5}{c}{Low--mass protostars}\\
\tableline
L1448--mm & $\sim 4.0 \times 10^{-7}$ & $\le 3.0 \times 10^{-9}$ & ... & ... & 1 \\
IRAS 16293-2422 & $1.0 \times 10^{-7}$ & $1.5 \times 10^{-10}$ & $3 \times 10^{-2}$ & $\le 2 \times 10^{-3}$ & 2 \\
               & $1.7 \times 10^{-7}$ & $8.0 \times 10^{-11}$ & $3.4\times 10^{-2}$ & $5.0\times 10^{-3}$ & 3 \\
               & ... & ... & $(9.2\pm2.6)\times 10^{-4}$ & ... & 4 \\
                         
NGC 1333-IRAS2A  & $8.0 \times 10^{-8}$ & $7.0 \times 10^{-10}$ & $\geq 1.0 \times 10^{-2}$ & $ \sim 7 \times 10^{-2}$ & 5\\
                                 & ... & ... & $(3.0-80)\times 10^{-3}$ & ... & 6\\
                                 & ... & ... & $1.0 \times 10^{-3}$ & ... & 7\\
                                 & ... & ... & $(7.4\pm2.1)\times 10^{-4}$ & ... & 8\\
NGC 1333-IRAS4B  & ... & ... & $\leq 6.0\times 10^{-4}$ & ... & 9 \\
                 & $1.0 \times 10^{-8}$ & $1.2 \times 10^{-10}$ & $(1.0-37) \times 10^{-4}$ & ... & 10\\
                                 & ... & ... & $(5.9\pm1.7)\times 10^{-4}$ & ... & 8\\
NGC 1333-IRAS 4A-NW              & ... & ... & $(19.1\pm5.4\times 10^{-4}$& ... & 8\\
                                 & $7.5 \times 10^{-9}$ & $1.2 \times 10^{-11}$ & $(4.0-30) \times 10^{-4}$ & ... & 10\\
                                  & ... & ... & $(5.0-30)\times 10^{-3}$ & ... & 6\\ 
\tableline
\multicolumn{5}{c}{Intermediate--mass protostar}\\
\tableline
NGC 7192 FIRS2& $4.0\times 10^{-8}$ & ... & ... & ... & 11\\
\tableline
\multicolumn{5}{c}{High--mass hot cores \tablenotemark{$\star$}}\\
\tableline
 G34.26+0.15 & $3.7\times 10^{-8}$ & $7.8\times 10^{-11}$ &  $3.7\times 10^{-4}$ & $3.1\times 10^{-3}$ & 12\\
             & $6.0\times 10^{-8}$ & $5.0\times 10^{-12}$ &  $3.0\times 10^{-4}$ & $(1.9-4.9)\times 10^{-4}$ & 13\\
             & $2.0\times 10^{-7}$ & $8.0\times 10^{-11}$ & $(3.5-7.5)\times 10^{-4}$ & $(1.0-2.2)\times 10^{-3}$ & 14\\
             & $4.6\times 10^{-9}$ & ... & $1.1\times 10^{-4}$ & ... & 15 \\
             & $ \sim 2.7\times 10^{-8}$ & ... & $4.0\times 10^{-4}$ & ... & 16 \\
 W51e$_{1}$/e$_{2}$  & $1.7\times 10^{-8}$ & $7.0\times 10^{-11}$ & $1.7\times 10^{-4}$ & $2.8\times 10^{-3}$ & 12\\
 W49N  & $3.0\times 10^{-9}$ & $1.0\times 10^{-10}$ & $3.0\times 10^{-5}$ & $1.8\times 10^{-3}$ & 12\\
                      & $1.5\times 10^{-9}$ & ... & $6.3\times 10^{-5}$ & ... & 15 \\
                      & $ \sim 2.2\times 10^{-8}$ & ... & $3.0\times 10^{-4}$ & ... & 16\\
 W 33A                & $2.0\times 10^{-7}$ & $1.0\times 10^{-8}$ & ... & ... & 17\\
 NGC 6334 I & $1.3\times 10^{-10}$ & ... & $2.1\times 10^{-4}$  & ... & 18\\
 W3$(OH)/(H_2O)$ & ... & ... & $(2.0-6.0)\times 10^{-4}$ & ... & 19\\
 AFGL 2591 & $1.0\times 10^{-7}$ & $2.0\times 10^{-8}$& ... & ... &17\\
 NGC 7538 IRS1  & $1.0\times 10^{-7}$ & $4.0\times 10^{-9}$& $5.0\times 10^{-4}$ & $ (4.0-400.0)\times 10^{-3}$ & 17\\
 Sgr B2(M) & $1.5\times 10^{-9}$ & $1.3\times 10^{-11}$& ... & ... & 20\\
           & $2.0\times 10^{-9}$ & ... &$1.8\times 10^{-4}$ & ... & 15\\
 Orion KL & $\sim 4.5\times 10^{-8}$ & ... & $3.0\times 10^{-3}$ & ... & 21 \\
\tableline
\tableline
\end{tabular}

\tablenotetext{\star}{The fractional abundances of water in the hot core:  $X_{in}(H_{2}O)$ $\sim1.0\times 10^{-4}$ (e.g.\citealt{Boonman03}, \citealt{Snell00}, \citealt{Herpin12}, \citealt{Emprechtinger13}) are taken in the inner HDO/H$_2$O ratio estimation.}
\tablerefs{
(1) Codella et al. 2010; (2) Parise et al. 2005; (3) Coutens et al. 2012; (4) Persson et al. 2014 on scale $\le$ 300 AU;
(5) Liu et al. 2011; (6) Taquet et al. 2013; (7) Visser et al. 2013; (8) Persson et al. 2014,on scale $\le$ 300 AU; (9) J\o rgensen et al. 2010, on scale $\le$ 50 AU; (10) Coutens et al. 2013;
(11) Fuente et al. 2012; (12) This work; (13) Liu et al. 2013; (14) Coutens et al. 2014; (15) Gensheimer et al. 1996; (16) Jacq et al. 1990; (17) van der Tak et al. 2006; (18) Emprechtinger et al. 2013; (19) Helmich et al. 1996; (20) Comito et al. 2010;  (21) Neill et al. 2013; (22) Lecuyer, C. et al. 1998.}
 
\end{table}
%
\acknowledgements
  {The author thanks  the referee very much for the highly constructive comments and suggestions. The author is grateful to N.\, Flagey for providing the reduced HIFI
 data of H$_2^{18}$O. M.K. would like to thank  Maryvonne Gerin and Darek Lis for fruitful discussions and 
  a careful reading of the manuscript. The work was carried out within the framework of the European
  Associated Laboratory "Astrophysics Poland-France" and also supported by the Science and High Education Ministry of Poland, grants 
  N20339/3334.
  This research is based on observations from
  the Caltech Submillimeter Observatory, which is operated by the
  California Institute of Technology under cooperative agreement with
  the National Science Foundation (AST-0838261).}

 \pagebreak

\bibliographystyle{aasjournal}
\bibliography{references.bib}

\end{document}